\documentclass[12pt,english]{article}
\usepackage[T1]{fontenc}
\usepackage[utf8]{inputenc}
\usepackage{geometry}
\usepackage{color}
\usepackage{babel}
\usepackage{float}
\usepackage{amsmath}
\usepackage{graphics}
\usepackage{graphicx}
\usepackage[authoryear]{natbib}
\usepackage{color}
\usepackage[dvips,dvipdfm,colorlinks=true,urlcolor=rblue,citecolor=rblue,linkcolor=rblue,bookmarks=true]{hyperref}
\definecolor{rblue}{cmyk}{0.71,0.53,0,0.12}
\usepackage{xfrac}   
\usepackage[hang]{caption}
\usepackage{bm} 
\usepackage{tikz}
\usetikzlibrary{shapes.geometric, arrows.meta, positioning}

\tikzstyle{startstop} = [rectangle, rounded corners, minimum width=3.5cm, minimum height=1cm,text centered, draw=black, fill=blue!20]
\tikzstyle{process} = [rectangle, minimum width=3.5cm, minimum height=1cm, text centered, draw=black, fill=gray!10]
\tikzstyle{arrow} = [thick,->,>=stealth]

\makeatletter

\numberwithin{figure}{section}

\usepackage{lmodern}

\usepackage{indentfirst}

\usepackage{pdflscape}

\usepackage{chngcntr}
\counterwithout{figure}{section}

\usepackage{booktabs}

\usepackage{afterpage}
\usepackage{multirow}

\makeatother

\begin{document}

\title{\textbf{Tracking the economy at high frequency}\\[1ex] \large\textit{Working Paper}}

\author{Yanina Baja\~{n}a-Villag\'omez\thanks{\raggedright Universidad Cat\'olica de Santiago de Guayaquil, Guayaquil, Ecuador. \mbox{Email: yanina.bajana@cu.ucsg.edu.ec}} \and Freddy Garc\'{i}a-Alb\'{a}n\thanks{\raggedright Universidad Ecotec, Km 13 1/2 V\'ia Samborond\'on, Samborond\'on, Guayas, Ecuador. \mbox{Email: fgarcia@ecotec.edu.ec}} \and Juan Jarr\'{i}n\thanks{\raggedright Ecuador. \mbox{Email: juanjarrin@outlook.es}} \and Ramon Villa-Cox\thanks{\raggedright ESPAE Business School, Escuela Superior Polit\'ecnica del Litoral, Guayaquil, Ecuador. \mbox{Email: ramavill@espol.edu.ec}}}

\date{July 2025}
\maketitle
\begin{abstract}
This paper develops a high-frequency economic indicator using a Bayesian Dynamic Factor Model estimated with mixed-frequency data. The model incorporates weekly, monthly, and quarterly official indicators, and allows for dynamic heterogeneity and stochastic volatility. To ensure temporal consistency and avoid irregular aggregation artifacts, we introduce a pseudo-week structure that harmonizes the timing of observations. Our framework integrates dispersed and asynchronous official statistics into a unified High-Frequency Economic Index (HFEI), enabling real-time economic monitoring even in environments characterized by severe data limitations. We apply this framework to construct a high-frequency indicator for Ecuador, a country where official data are sparse and highly asynchronous, and compute pseudo-weekly recession probabilities using a time-varying mean regime-switching model fitted to the resulting index.\\

\end{abstract}
{\let\thefootnote\relax\footnotetext{The views expressed in this paper are solely the responsibility of the authors and should not be interpreted as reflecting the views of their affiliated institutions.}}

\newpage
\section{Introduction}\label{Intro}

Timely and accurate assessment of current economic conditions is a central task in macroeconomic analysis. Such assessments are crucial not only for guiding public policy choices, but also for informing a wide range of decisions in the private sector. Businesses rely on up-to-date economic signals to plan production, manage inventories and adjust pricing strategies. Financial institutions depend on real-time information to assess credit risk and manage liquidity. Investors, both domestic and foreign, monitor short-term indicators to gauge macroeconomic trends and adjust portfolio allocations accordingly. However, producing reliable real-time estimates of economic activity remains challenging, particularly in emerging economies, where institutional constraints and data limitations are more severe. 

Ecuador offers a particularly illustrative case of this problem. While a variety of official indicators are produced at monthly, and quarterly frequencies, the country lacks a coherent, integrated framework for high-frequency monitoring of the real economy. The Central Bank of Ecuador calculates leading and coincident indicators, but these are available only at quarterly frequency and with significant lags (see \citet{BCE2021Nota87} and \citet{BCE2021Nota86}). High-frequency administrative data do exist, but they are often dispersed across public institutions, difficult to access in structured form, and not readily usable for economic tracking without considerable preprocessing.

This institutional gap leaves policymakers, analysts and private agents with limited tools to assess current macroeconomic conditions.

A few non-governmental initiatives have emerged to address the absence of timely economic indicators. For example, the \textit{Encuesta de Expertos de la Econom\'ia del Ecuador} publishes monthly forecasts for current and year-end macroeconomic variables based on a panel of professional economists, following a methodology similar to the Survey of Professional Forecasters in the United States. This survey provides valuable insights into market expectations and serves as a benchmark for anticipated trends in inflation, output, and other key variables. More recently, \citet{romero2025uindex} developed an index of economic uncertainty for Ecuador using text analysis of news articles. The uncertainty index is updated on a weekly or monthly basis and offers a forward-looking perspective on the informational environment surrounding economic policy and performance. While these initiatives represent important steps toward improving real-time economic monitoring in Ecuador, they are focused on expectations and perceptions rather than on measuring the underlying evolution of economic activity.

This paper contributes to closing that gap by constructing a high-frequency indicator of aggregate economic activity for Ecuador. To do so, we assemble a novel dataset composed entirely of official sources, integrating variables at weekly, monthly, and quarterly frequencies.

A large and growing body of literature is devoted to tracking current economic conditions in real time. Much of this work focuses on models based on monthly and quarterly data, with the aim of constructing coincident indices or nowcasting specific macroeconomic aggregates such as GDP. Pioneering studies in this area include \citet{stock1988probability}, \citet{stock1989new}, \citet{mariano2003new}, and \citet{camacho2010introducing}. As noted by \citet{baumeister2024tracking}, interest in higher-frequency indicators expanded significantly in the wake of the COVID-19 pandemic, as economists and policymakers increasingly turned to such tools to monitor the rapid deterioration in economic conditions driven by the public health crisis and the associated containment measures.

A relevant contribution along these lines is the Weekly Economic Index (WEI) of \citet{lewis2022measuring}, who relies exclusively on weekly time series and scales the resulting index to match the four-quarter growth rate of real GDP. Building on the findings of \citet{aruoba2009real}, who show that incorporating weekly data improves the accuracy of a real activity factor, \citet{baumeister2024tracking} develop a mixed-frequency factor model that blends quarterly, monthly and weekly indicators to estimate economic conditions across the 50 U.S. states. \citet{chan2023high} present another recent contribution, combining data at multiple frequencies within a large Bayesian VAR framework to produce weekly GDP estimates for the U.S. They also suggest that pure weekly indicators, such as the WEI, may primarily reflect activity in specific sectors of the economy, whereas mixed-frequency models like their own and the business conditions index of \citet{aruoba2009real} provide a more comprehensive picture by incorporating information from official GDP data.

Our empirical approach is closely related to that of \citet{baumeister2024tracking}. We build a Bayesian dynamic factor model (BDFM) that extracts a latent factor summarizing the co-movements across a large set of variables observed at different frequencies. This factor is interpreted as a weekly indicator of economic activity and serves as a timely and interpretable proxy for the underlying state of the business cycle. However, our model differs from that of \citet{baumeister2024tracking} in three key respects. First, we rely on the use of pseudo-weeks instead of standard calendar weeks. As will be explained in more detail later, this approach mitigates mechanical issues associated with the fact that some months and quarters contain a different number of weeks. Second, we extend the model by allowing for heterogeneous dynamics in the factor loadings, as proposed by \citet{antolin2024advances}. Third, we introduce richer dynamics by incorporating stochastic volatility, which is particularly relevant in the context of the COVID-19 pandemic, as emphasized by \citet{almuzara2023new}, \citet{lenza2022estimate}, \citet{hartwig2024bayesian}, and \citet{carriero2024addressing}.

The high-frequency indicator presented in this paper provides a new tool for real-time economic monitoring in Ecuador. It can be updated regularly as new data become available. While the empirical focus is on Ecuador, the modeling choices and data handling procedures are general enough to be adapted to other contexts where high-frequency measurement of economic activity is needed.

The remainder of the paper is structured as follows. Section \ref{dataset} describes the construction of the mixed-frequency dataset. Section \ref{econframe} presents the specification of the BDFM and the estimation procedure. Section \ref{wei} discusses the resulting High-Frequency Economic Index (HFEI), including its interpretation and the estimation of recession probabilities. Section \ref{complexity} examines the trade-off between structural complexity and empirical performance across alternative model specifications. Section \ref{conclusions} concludes.

\section{The mixed-frequency dataset}\label{dataset}
We construct a novel dataset of Ecuadorian macroeconomic indicators sampled at quarterly, monthly, and weekly frequencies. The following subsections describe the rationale for the variable selection and key data processing steps.

\subsection{Variable selection}
We use a sample of seventeen series ranging from 2004 to the fourth quarter of 2024. All the series are updated to the end of the period. However, some of them do not contain data for some years at the beginning of the sample. This implies an unbalanced dataset with missing observations caused by the mixed frequency, but also for the non-existence of data at the beginning of the sample. Table \ref{tab:desc_variables} lists all the variables used, their sample frequency and publication lag.

\begin{table}[htbp]
\centering
\caption{Data series used for the empirical analysis}
\label{tab:desc_variables}
\begin{tabular}{llll}
\hline
\textbf{Variable} & \textbf{Group} & \textbf{First observation} & \textbf{Lag} \\
\hline
\multicolumn{4}{l}{\textit{Quarterly indicators}} \\
\hspace{1em} Real GDP & Real activity & 2004Q1 & 105 \\
\hspace{1em} Real Household Consumption & Real activity & 2004Q1 & 105 \\
\hspace{1em} Credit supply index & Expectations & 2019Q1 & 40 \\
\hspace{1em} Credit demand index & Expectations & 2019Q1 & 40 \\
\addlinespace
\multicolumn{4}{l}{\textit{Monthly indicators}} \\
\hspace{1em} National Sales & Real activity & 2012M1 & 45 \\
\hspace{1em} Vehicle Sales & Real activity & 2008M1 & 15 \\
\hspace{1em} Economic Expectations Index & Expectations & 2011M2 & 30 \\
\hspace{1em} Adequate Employment & Labor market & 2015M1 & 23 \\
\hspace{1em} Oil Production & Real activity & 2004M1 & 45 \\
\addlinespace
\multicolumn{4}{l}{\textit{Weekly indicators}} \\
\hspace{1em} Firm creation & Expectations & 2004W1 & 7 \\
\hspace{1em} Private Credit & Money and credit & 2012W2 & 11 \\
\hspace{1em} Private Deposits & Money and credit & 2012W2 & 11 \\
\hspace{1em} Non-oil Exports & Trade & 2012W1 & 45 \\
\hspace{1em} Oil Exports & Trade & 2014W1 & 45 \\
\hspace{1em} Imports of Consumer Goods & Trade & 2004W1 & 45 \\
\hspace{1em} Imports of Capital Goods & Trade & 2004W1 & 45 \\
\hspace{1em} Imports of Raw Materials & Trade & 2004W1 & 45 \\
\hline
\end{tabular}

\vspace{0.5em}
\begin{minipage}{0.95\textwidth}
\small \textit{Note:} The last column shows the average publication lag, i.e., the number of days elapsed from the end of the period that the data point refers to until its publication by the statistical agency.
\end{minipage}
\end{table}

We use four quarterly variables: real GDP, household consumption, and two survey-based credit indexes. GDP and consumption are published with a significant delay, typically three to four months after the end of the reference quarter. Timely quarterly indicators based on hard data beyond the GDP components of the national accounts are not available for Ecuador. For instance, variables such as Gross Domestic Income or labor costs are not reported at a quarterly frequency. Other potentially relevant series, such as Foreign Direct Investment and remittances, are available quarterly but with even longer publication lags.

In addition to these, we include two expectation-based indicators of credit: a credit supply index and a credit demand index. Both are derived from surveys conducted by the Central Bank of Ecuador with more than 300 financial and banking institutions. The supply index reflects changes in credit approval standards used by banks, while the demand index captures variation in new loan applications. These indexes track changes in the underlying qualitative measures and are correlated with annual credit growth and other indicators of the business cycle.

Regarding the monthly variables, we use five macroeconomic time series: national  sales, vehicle sales, the Economic Expectations Index, adequate employment and oil production. National sales and vehicles sales are administrative data collected by the tax authority. While national sales are reported in dollars, vehicles sales are reported in units. The Economic Expectations Index is an indicator published by the Central Bank of Ecuador based on surveys conducted with the most representative companies in Ecuador.

For the adequate employment series, we use official labor data from the National Institute of Statistics and Censuses (INEC), which is published monthly. However, prior to September 2020, the data were only available for March, June, September, and December. To obtain a consistent monthly series, we interpolate the missing months using information from social security records. A full description of this procedure is provided in Appendix \ref{app:data}.

Oil production is included among the monthly indicators due to its central role in shaping Ecuador's economic dynamics. The economy's dependence on oil revenues makes it particularly sensitive to fluctuations in this sector. As shown by \citet{garcia2021good}, oil revenue shocks account for a large share of output volatility in Ecuador, often exceeding the contribution of fiscal policy. Including oil-related variables in high-frequency models is therefore essential for capturing the transmission of these shocks to real activity.

It is possible to disaggregate most of the monthly data to a more granular level, but we prefer to use only headline indicators. As explained by \citet{boivin2006more}, \citet{banbura2013now} and \citet{alvarez2012finite}, the presence of strong correlation in the idiosyncratic components of disaggregated series of the same category will be a source of misspecification that can worsen the in-sample and out-of-sample performance of the model.

We exclude price indexes from the set of monthly variables, as we believe they fail to capture meaningful fluctuations in Ecuador's business cycle. This may be related to the country's condition as a fully dollarized economy without conventional monetary policy instruments. Moreover, the literature on nowcasting GDP has documented that price measures often provide limited information for tracking real activity (see \cite{sargent1977business}, \cite{BRAGOLI2017}).

Regarding the weekly series, we use three subsets of variables, comprising a total of eight indicators. The first group includes five trade variables related to exports and imports. The second consists of money and credit variables, specifically private sector credit and deposits, and the third includes a variable capturing firm creation. We classify the variables in this manner to subsequently analyze the role of trade and money indicators in explaining high-frequency fluctuations of the business cycle. These variables are commonly used, either in formal models or for trend monitoring purposes, at monthly or quarterly frequencies. To the best of our knowledge, we are the first to construct a comprehensive dataset of these variables at a weekly frequency and to incorporate them into an econometric model for the Ecuadorian economy.

The weekly trade dataset consists of oil and non-oil exports, as well as imports of consumer goods, capital goods, and raw materials. Export series are included in U.S. dollars, while import series are expressed in metric tons, Although these data are available at a weekly or even daily frequency, they are only published on a monthly basis. In this sense, the higher frequency does not provide additional value in terms of timing compared to their monthly counterparts. However, they do help characterize high-frequency fluctuations without the need to rely on additional latent variables, which would increase the uncertainty of the estimates.

The weekly money and credit dataset includes private sector credit and deposit balances, obtained from the Central Bank of Ecuador. These monetary aggregates reflect lending activity and liquidity conditions in the economy, and offer valuable high-frequency signals of financial behavior and demand-side pressures. Their inclusion in the model helps capture short-term macro-financial dynamics, particularly during periods of heightened uncertainty or regulatory intervention.

The firm creation variable captures the weekly number of newly registered firms, based on administrative records from the Superintendence of Companies, Securities and Insurance. While it reflects real sector dynamics, it also carries an expectations component. The decision to formalize a business typically involves forward-looking assessments of market conditions, credit availability, and policy stability. As such, new firm registrations offer a high-frequency signal of both current economic activity and business sentiment.

\subsection{Pseudo-weeks}\label{pseudo-weeks}
Apart from the quarterly and monthly series, the remaining data is originally reported at a daily frequency. Some variables are officially labeled as weekly by the original source. However, in practice, these correspond to values recorded on the final working day of the week which may vary due to holidays or other operational factors. Therefore, we treat them as daily observations with irregular spacing.

We aggregate the daily data into pseudo-weekly frequency. A pseudo-week is an artificial temporal unit that divides each month into four equal periods, ensuring intra-month consistency while allowing for weekly aggregation. By construction, this also implies intra-quarter and year consistency meaning that each quarter and year always have twelve and forty-eight weeks, respectively. The pseudo-week structure is defined as follows:

\begin{table}[H]
\centering
\caption{Definition of Pseudo-Weeks Within a Month}
\begin{tabular}{cc}
\hline
\textbf{Day of the Month} & \textbf{Pseudo-Week} \\
\hline
1--7   & Week 1 \\
8--14  & Week 2 \\
15--21 & Week 3 \\
22--End of Month & Week 4 \\
\hline
\end{tabular}
\label{tab:pseudo_weeks}
\end{table}

Using pseudo-weeks compared to using standard calendar weeks has some advantages. First, the matrix of coefficients of the observation equation in the state-space representation of the model is time-invariant with fixed size. As illustrated by \citet{baumeister2024tracking} and \citet{chan2023high} the use of calendar weeks in mixed-frequency models involves time-varying and size-changing matrices generating irregular missing data patterns. Second, the year-over-year growth rates are more likely to remove yearly seasonal patterns in the case of pseudo-weeks. Third, pseudo-weeks allow for cleaner comparisons of activity across similar periods making the analysis of intra-month effects more transparent.

In our model, we use stock and flow variables at a weekly frequency, so their aggregation over a pseudo-week follows different rules. Stock variables reflect a level at a specific point in time, so their pseudo-weekly values are obtained by taking the mean within each pseudo-week. In contrast, flow variables measure activity accumulated over time, and their pseudo-weekly values are obtained by summing the corresponding daily observations.

\subsection{An overview of the data and some stylized facts about the Ecuadorian economy}

Ecuador, as an oil-exporting economy, has been particularly vulnerable to external shocks and abrupt changes in global conditions. Over the past two decades, the country has experienced four major episodes of economic contraction or deceleration: the 2008-09 global financial crisis and the associated collapse in oil prices; the 2015-16 downturn, also triggered by a sharp fall in oil revenues; the 2020 crisis caused by the COVID-19 pandemic; and the more recent slowdown in 2024. While the drivers of the 2024 episode are less clearly defined, it appears to be linked to rising political and economic uncertainty, coupled with severe liquidity constraints in the public sector. The following analysis uses high-frequency indicators to characterize the behavior of key macroeconomic variables across these episodes and to highlight structural features of the Ecuadorian economy.

\begin{figure}
\centering
\caption{\label{fig:credit_deposits_sales} Private Credit, Deposits, and Sales}
\includegraphics[width=\textwidth]{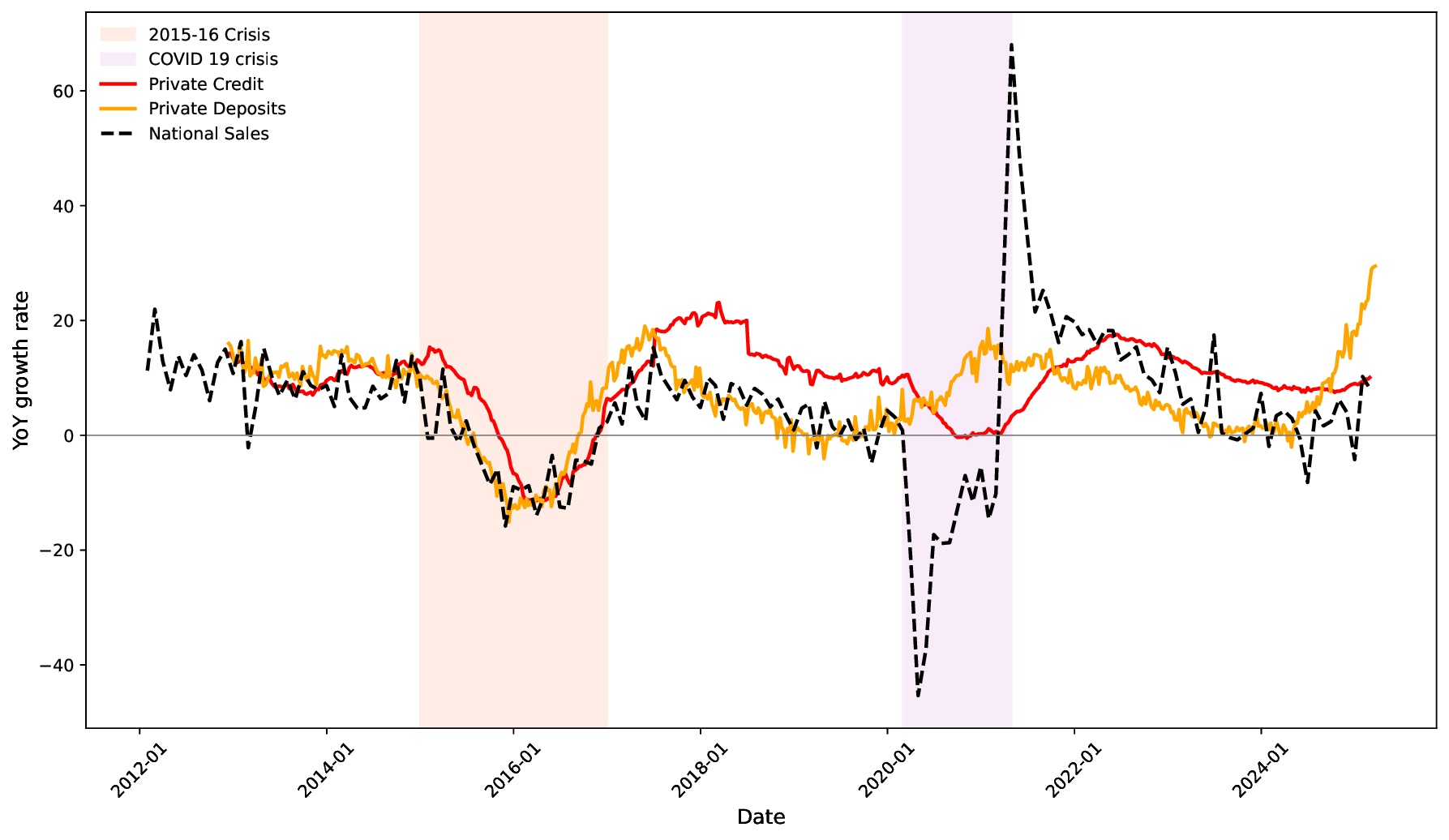}
\caption*{\tiny{\emph{Note:} The figure displays the year-over-year growth rate in private credit (black), private deposits (orange), and national sales (red dashed). The shaded areas indicate the 2015--16 economic crisis (light salmon) and the COVID-19 crisis (light gray).}}
\end{figure}

Monetary aggregates such as credit and deposits typically co-move with the business cycle, expanding during periods of growth and contracting during recessions. However, the COVID-19 shock in Ecuador displayed a notable deviation from this pattern (see Figure \ref{fig:credit_deposits_sales}). Despite a sharp contraction in real activity, private credit continued to grow. Weekly data show only a brief stagnation in lending during the strict lockdown period, with no sustained decline. This atypical behavior can be partly explained by extraordinary deferral mechanisms implemented by financial institutions in response to the pandemic. These measures allowed clients to restructure, refinance, or postpone their credit obligations. At the same time, deposits increased steadily, likely reflecting a rise in precautionary savings and a reduction in consumption during the lockdown.

The divergence between the continued growth in credit and deposits and the sharp contraction in real activity indicators underscores the unusual monetary dynamics observed during the pandemic. While monetary aggregates appeared resilient, other indicators painted a much more severe picture. For instance, monthly national sales fell by nearly $40\%$ during the peak of the lockdown. By comparison, the 2015-16 recession exhibited a more synchronized decline across credit, deposits, and sales. Moreover, the disconnection between real activity and deposits appears to persist even in the post-pandemic period.

\begin{figure}
\centering
\caption{\label{fig:imp_weekly} Import dynamics by type of good}
\includegraphics[width=\textwidth]{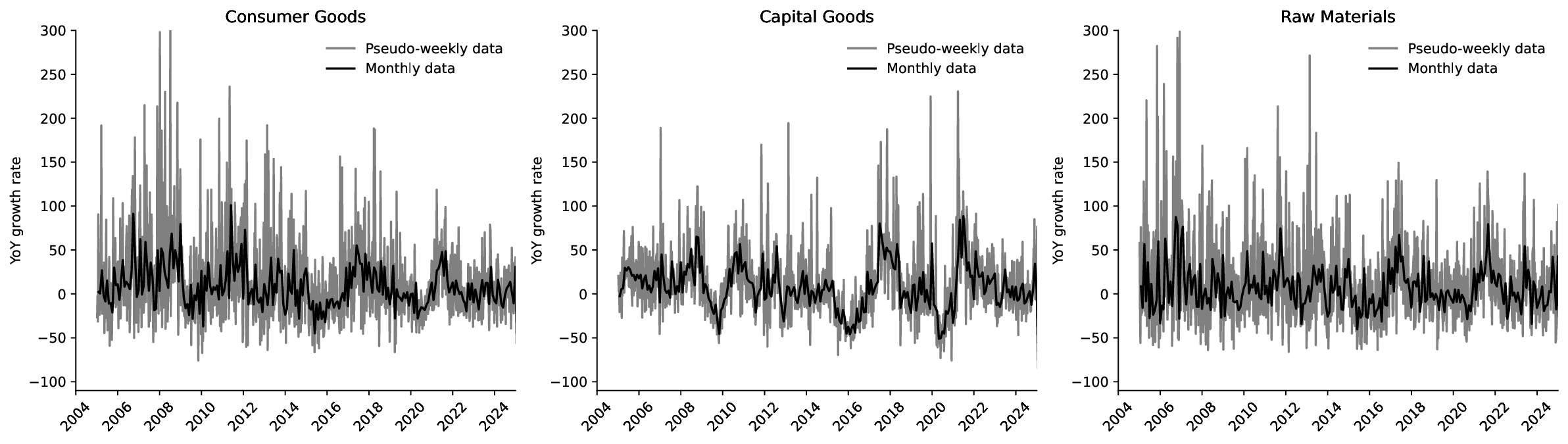}
\caption*{\tiny{\emph{Note:} The figure displays the year-over-year growth rate of metrics tons in weekly imports of consumer goods, capital goods, and raw materials. Each panel includes both the full pseudo-weekly series (colored line) and the monthly series (black line) consisting of a representative weekly observation per month.}}
\end{figure}

On the other hand, high-frequency data on Ecuador's trade offer valuable insights into how different sectors move over the business cycle. Disaggregated import series show that imports of consumer and capital goods are particularly sensitive to crisis episodes. Sharp contractions are evident during the 2008-09 global financial crisis, the 2015-16 fiscal and oil shock, and the 2020 COVID-19 pandemic. These patterns are visible in the year-over-year growth rates shown in Figure \ref{fig:imp_weekly}.

In contrast, imports of raw materials exhibit more volatile behavior. Their movements are less tightly linked to domestic cycles, likely reflecting demand from export-oriented sectors such as shrimp or banana production, which remained active even during lockdowns or fiscal adjustments.

Figure \ref{fig:imp_weekly} also highlights the distinction between high-frequency (pseudo-weekly) and lower-frequency (monthly) import data. Many of the sharpest contractions, especially those occurring at the onset of crises, unfold over short periods that are clearly captured by weekly indicators but are blurred or even missed in monthly data. In the case of Ecuador, where shocks tend to materialize abruptly and official statistics are published with lags, relying solely on monthly indicators may underestimate both the timing and severity of real activity disruptions.

\begin{figure}[H]
\centering
\caption{\label{fig:exports_oil_non_oil} Oil and Non-Oil exports}
\includegraphics[width=\textwidth]{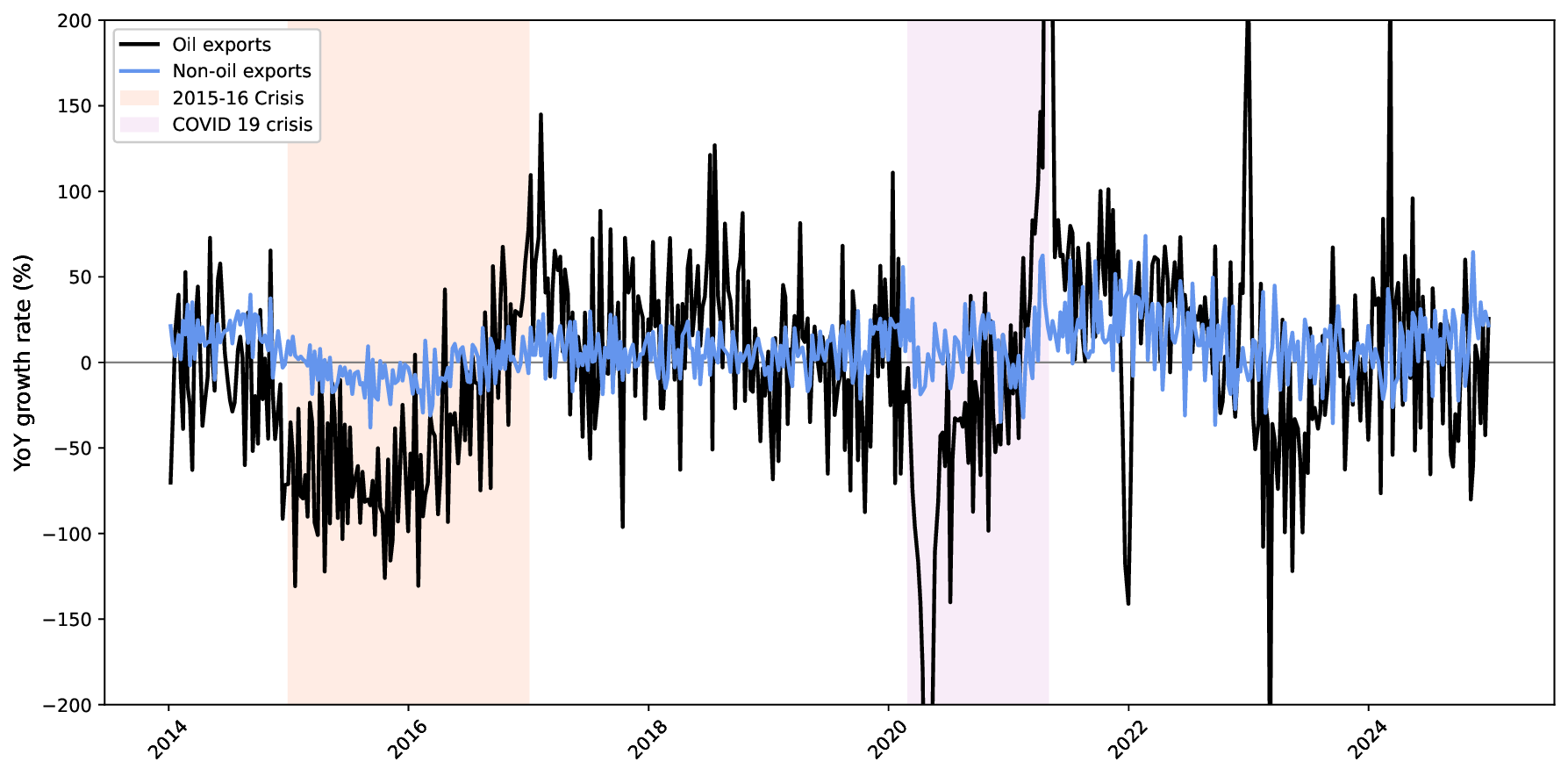}
\caption*{\tiny{\emph{Note:} The figure shows the year-over-year growth rate FOB oil exports (black) and non-oil exports (light blue). The vertical axis has been limited to $\pm$200\% for visualization purposes. The maximum value reaches 569\% and the minimum value reaches -614\%}} 
\end{figure}

Figure \ref{fig:exports_oil_non_oil} shows that Ecuador's oil exports experience sharp declines during recessionary periods and strong recoveries when activity rebounds. However, this pattern does not reflect a response to the domestic business cycle. Instead, it illustrates the extent to which the Ecuadorian economy is exposed to fluctuations in international oil prices, over which it has no control. As a price-taking exporter, Ecuador's oil revenues are driven by external shocks, which in turn shape the domestic cycle. In contrast, non-oil exports, mainly agricultural and fishery products, show more muted responses during crises, likely due to more stable external demand or contractual rigidities that help cushion downturns. This contrast was especially pronounced during the COVID-19 crisis, when oil exports contracted much more sharply than non-oil exports. As emphasized by \citet{garcia2021good}, oil revenue shocks have historically been the most important force moving Ecuadorian output above or below trend, highlighting the country's structural dependence on oil.\footnote{Oil price shocks are transmitted to the Ecuadorian economy primarily through the fiscal channel, given the centrality of oil revenues in public finances, and through the trade balance, as oil remains one of the country's main export products.}

While weekly trade variables are highly volatile and responsive to short-term fluctuations, weekly credit and deposit series tend to display smoother dynamics, reflecting structural and institutional frictions in the financial system that dampen short-run variation.

\section{Econometric framework}\label{econframe}
We implement a dynamic factor model that accommodates mixed-frequency data, dynamic heterogeneity and time-varying volatility. The model is cast in state-space form and estimated using Bayesian methods.

\subsection{The dynamic factor model}

Let $\mathbf{y}_t$ be an $n\times1$ vector of macroeconomic time series. We assume that there is a single latent weekly cyclical factor, $f_t$, that is common to all the variables in the system and captures most of the comovement between them. Formally,

\begin{align}
\mathbf{y}_{t}=\mathbf{\Lambda(L)}f_{t}+\mathbf{u}_{t} \label{eq:model1}
\end{align}

where $\mathbf{\Lambda(L)}$ is a matrix polynomial of order $s$ in the lag operator containing the loadings on the contemporaneous and lagged common factors, and $\mathbf{u}_t$ is a vector of idiosyncratic components. Following \citet{dagostino2016} and \citet{antolin2024advances}, when observables load on lagged values of the factor in the measurement equation $(s > 0)$, the model is commonly referred to as a dynamic factor model with dynamic heterogeneity or lead-lag dynamics.

We assume that the factor and the idiosyncratic errors follow AR($p_f$) and AR($p_q$) processes, respectively. As is commonly assumed, the idiosyncratic components are cross-sectionally orthogonal and uncorrelated with the common factor at all horizons, so the only source of correlation is given by the loadings.

Finally, we model the log of stochastic volatilities as driftless random walks. So the laws of motion of the latent variables are given by:

\begin{align}
(1-\phi(L))f_{t} &= \sigma_{\varepsilon_{t}}\varepsilon_{t}, &  \varepsilon_{t}\overset{\text{iid}}\sim N(0,1)\\
\ln\sigma_{\varepsilon_{t}}^2&=\ln\sigma_{\varepsilon_{t-1}}^2+\upsilon_{\varepsilon_{t}}, & \upsilon_{\varepsilon_{t}}\overset{\text{iid}}\sim N(0,\omega_{\varepsilon}^{2})
\end{align}
\begin{align}
(1-\rho_{i}(L))u_{i,t} &= \sigma_{\eta_{i,t}}\eta_{i,t}, & \eta_{i,t}\overset{\text{iid}}\sim N(0,1), & & i=1,\ldots,n \label{eq:model2} \\
\ln\sigma_{\eta_{i,t}}^2&=\ln\sigma_{\eta_{i,t-1}}^2+\upsilon_{\eta_{i,t}}, & \upsilon_{\eta_{i,t}}\overset{\text{iid}}\sim N(0,\omega_{\eta_{i}}^{2}), & & i=1,\ldots,n
\end{align}

where $\phi(L)$ and $\rho_i(L)$ denote polynomials in the lag operator of orders $p_f$ and $p_q$.

\subsection{Mixed frequencies}

There are $n_{q}$ quarterly variables, $n_{m}$ monthly variables and $n_{w}$ weekly variables such that $n=n_w+n_m+n_q$. We model all variables in weekly frequency, and the missing observations of the monthly or quarterly variables are linked to their corresponding observed values via inter-temporal constraints similar to those in \citet{mariano2003new} and \citet{camacho2010introducing}. Specifically, any observable monthly variable can be expressed as four times the arithmetic mean of its unobservable weekly counterpart:

\begin{equation}
Y_{m_{t}}=4\left(\frac{Y_{w_{t}}+Y_{w_{t-1}}+Y_{w_{t-2}}+Y_{w_{t-3}}}{4}\right)
\end{equation}

After approximating the arithmetic mean by the geometric mean and taking logs, we obtain:

\begin{equation}
\ln{Y_{m_{t}}}=\ln{4}+\frac{1}{4}\left(\ln{Y_{w_{t}}}+\ln{Y_{w_{t-1}}}+\ln{Y_{w_{t-2}}}+\ln{Y_{w_{t-3}}}\right) \label{eq:log}
\end{equation}

We focus on the year-over-year growth rate, such that we are interested in $\ln{Y_{m_{t}}}-\ln{Y_{m_{t-12}}}$. It is straightforward to show that using the approximation in (\ref{eq:log}):

\begin{equation}
\ln{Y_{m_{t}}}-\ln{Y_{m_{t-12}}}=\frac{1}{4}\left(y_{w_t}+y_{w_{t-1}}+y_{w_{t-2}}+y_{w_{t-3}}\right),
\end{equation}

where $y_{w_t} = \ln{Y_{w_t}}-\ln{Y_{w_{t-48}}} $ the weekly year-over-year growth rate.

A similar structure applies to quarterly variables. We can express the year-over-year growth rate of variables at quarterly frequency in terms of the year-over-year growth rate of the weekly latent variables as follow:

\begin{equation}
\ln{Y_{q_{t}}}-\ln{Y_{q_{t-4}}}=\frac{1}{12}\left(y_{w_t}+y_{w_{t-1}}+\ldots+ y_{w_{t-11}}\right)
\end{equation}

By combining these aggregation rules with (\ref{eq:model1}) and (\ref{eq:model2}), we can express the observables monthly and quarterly variables as functions of the factor and the idiosyncratic error terms. This allows for a parsimonious state-space representation of the model that naturally accommodates mixed-frequency observations.

\subsection{Priors and model settings}
The state-space form of the model is given by:

\begin{align}
\mathbf{\tilde{y}}_{t} &= \mathbf{H}\bm{\xi}_{t} \label{state1}\\
\bm{\xi}_{t}& =\mathbf{F}\bm{\xi}_{t-1}+\mathbf{R}_{t}\mathbf{v}_{t}\label{state2}
\end{align}

where the state vector contains current and lagged values of the factor and idiosyncratic errors. Appendix \ref{app:model} provides more detail about the construction of the state-space form.

We estimate the model within a Bayesian framework, using its state-space representation to jointly infer the latent variables, time-varying volatilities, and missing observations via the Kalman filter and simulation smoother.

We follow \citet{antolin2024advances} and use priors that shrinks the parameters towards a more parsimonious specification. We impose Minnesota-type priors on the autoregressive coefficients in both $\phi(L)$ and $\rho_{i}(L)$. For the common factor, the prior mean is set to 0.9 on the first lag and zero on all higher-order lags, reflecting a belief that the factor captures a highly persistent process. For the idiosyncratic components, the priors on all autoregressive coefficients are centered at zero, encouraging shrinkage toward a model with no serial correlation. As in the Bayesian VAR literature the variances of these priors are $\frac{\gamma}{h^2}$ where $\gamma$ is set to 0.2 and $h$ is equal to the lag number of each coefficient.

For the variances $\omega_{\eta_{i}}^{2}$ and $\omega_{\varepsilon}^{2}$ we set an inverse gamma prior with one degree of freedom and scale equal to 0.0001. This choice favors small values, shrinking the volatilities toward zero and expressing a preference for homoskedastic error terms.

We set the number of lags in the autoregressive polynomials $\phi(L)$ and $\rho_i(L)$ to $p_f = 2$ and $p_q = 3$, respectively. These values are close to those used by \citet{baumeister2024tracking} in the context of the U.S. economy ($p_f = 4$, $p_q = 4$). Model comparison is conducted across alternative configurations of stochastic volatility and dynamic heterogeneity in the measurement equation.

We evaluate eight model specifications that vary along two dimensions: the degree of dynamic heterogeneity and the structure of stochastic volatility. Following \citet{antolin2024advances}, who show that $s = 1$ performs better than higher orders in U.S. data, we restrict our attention to $s \in {0,1}$.

Regarding stochastic volatility, most of the literature either assumes homoskedastic errors or allows for time-varying volatility in both the factor and the idiosyncratic components simultaneously. In contrast, we explore the possibility of introducing stochastic volatility selectively, either in the factor innovations or in the idiosyncratic errors, but not necessarily in both. Specifically, we evaluate four volatility configurations: homoskedasticity, volatility in the factor only, volatility in the idiosyncratic components only, and volatility in both.

While much of the literature focuses on predictive accuracy, especially for nowcasting, we instead assess model performance based on the ability of the latent factor to summarize co-movements in the observed variables and to capture stylized facts of the Ecuadorian economy. We adopt a dual evaluation strategy, combining a narrative assessment of model interpretability with the Deviance Information Criterion (DIC) proposed by \citet{spiegelhalter2002bayesian}, a widely used Bayesian criterion for in-sample model comparison.

The DIC balances model fit and complexity and serves as a Bayesian alternative to classical information criteria. Following \citet{spiegelhalter2002bayesian} and the refinement by \citet{chan2016observed}, the deviance is defined as
\[
D(\bm{\theta}) = -2 \log f(\mathbf{y} \mid \bm{\theta}) + 2 \log h(\mathbf{y}),
\]
where \( f(\mathbf{y} \mid \bm{\theta}) \) is the likelihood function and \( h(\mathbf{y}) \) is solely a function of the data, often assumed constant for simplicity. The effective number of parameters, denoted \( p_D \), measures model complexity and is given by
\[
p_D = \overline{D(\bm{\theta})} - D(\widetilde{\bm{\theta}}),
\]
where \( \overline{D(\bm{\theta})} \) is the posterior mean deviation and \( \widetilde{\bm{\theta}} \) is a point estimator of the parameters, typically the posterior mean. Thus, the DIC is computed as
\[
\text{DIC} = \overline{D(\bm{\theta})} + p_D.
\]
Under the assumption that \( h(\mathbf{y}) = 1 \), the DIC simplifies to
\[
\text{DIC} = -4\, E_\theta \bigl[ \log f(\mathbf{y} \mid \bm{\theta}) \,\big|\, \mathbf{y} \bigr] 
+ 2 \log f\bigl(\mathbf{y} \mid \widetilde{\bm{\theta}}\bigr),
\]
where the first term denotes the average log-likelihood over posterior draws.\footnote{When there are latent variables, $\mathbf{Z}$, in the model, it is possible to define at least three different measures of DIC. What we use here is the conditional DIC, which is based on the conditional likelihood $f(\mathbf{y} \mid \mathbf{Z}, \bm{\theta})$, using the terminology of \citet{celeux2006deviance} and \citet{chan2016fast}.} Lower DIC values indicate a better balance between model fit and parsimony.

\subsection{Estimation algorithm}
Let $\bm{\theta}\equiv \{ \bm{\Lambda}, \bm{\phi}, \bm{\rho}, \omega_{\eta_i}, \omega_{\varepsilon}\}$ be the vector containing all the parameters of the model. The latent components of the model, collected in the state vector $\bm{\xi}_t$ are $\{f_t,u_{i,t}\}_{t=1}^T$, and the sequences of stochastic volatilities are $\{\sigma_{\varepsilon}, \sigma_{\eta_i}\}_{t=1}^T$. Let $\mathbf{\tilde{y}}$ represent the full dataset of economic variables observed at quarterly, monthly, and weekly frequencies.

Bayesian estimation is carried out using the Gibbs sampler, which consists of the following steps:\footnote{For brevity, we present the algorithm for the specification with both dynamic heterogeneity and full stochastic volatility, which nests all other models considered. The derivation of the Gibbs steps for the restricted models is straightforward.}

\begin{enumerate}
    \setcounter{enumi}{-1} 
    \item \textit{Initialization}
    
    The model parameters and stochastic volatilities are initialized at arbitrary starting values $\bm{\theta}^{0}$ and $\{\sigma_{\varepsilon}^0, \sigma_{\eta_i}^0\}_{t=1}^T$.
    
    \item \textit{Draw the common factor and idiosyncratic components conditional on the data, the model parameters and stochastic volatilities}
    
    Obtain a draw $\{f_t\}_{t=1}^T$ and $\{u_{i,t}\}_{t=1}^T$ from $p(\{f_t,u_{i,t}\}_{t=1}^T|\bm{\theta}^{j-1},\{\sigma_{\varepsilon}^{j-1}, \sigma_{\eta_i}^{j-1}\}_{t=1}^T,\mathbf{\tilde{y}})$
    
    Using the state-space representation equations (\ref{state1}) and (\ref{state2}), generate a draw of the entire state vector $\bm{\xi}_t$ from the simulation smoother using the algorithm of \citet{durbin2002simple}.
    
    \item \textit{Draw the factor loadings}

    Obtain a draw $\bm{\Lambda}^j$ from $p(\bm{\Lambda}|\bm{\rho}^{j-1},\{f_t^j, \sigma_{\eta_i}^{j-1}\}_{t=1}^T,\mathbf{\tilde{y}})$

    Conditional on the draw of the common factor $f_t^j$ the measurement equations reduce to $n$ independent linear regressions with heteroskedastic and serially correlated residuals. Each element of $\bm{\Lambda}^j$ can be estimated by GLS, equation by equation, using $\bm{\rho}^{j-1}$ and ${\sigma_{\eta_i}^{j-1}}$. To ensure model identification, we normalize the contemporaneous loading of the GDP equation to 1.

    \item \textit{Draw the serial correlation coefficients of the idiosyncratic components}

    Obtain a draw $\bm{\rho}^j$ from $p(\bm{\rho}|\bm{\Lambda}^{j-1},\{f_t^j, \sigma_{\eta_i}^{j-1}\}_{t=1}^T,\mathbf{\tilde{y}})$

    Given the idiosyncratic components and the sequence for their stochastic volatilities, obtain $\bm{\rho}^j$ from simple autoregressions with heteroskedasticity. The conditional posterior is a Normal distribution. Draws which imply non-stationary behavior are rejected.
    
    \item \textit{Draw the factor autoregressive parameters}

    Obtain a draw $\bm{\phi}$ from $p(\bm{\phi}|\{f_t^j,\sigma_{\varepsilon_t}^{j-1}\}_{t=1}^T)$

    The autoregressive coefficient can be sampled conditional on the factor $\{f_t^j\}_{t=1}^T$ and its stochastic volatility $\{\sigma_{\varepsilon_t}^{j-1}\}_{t=1}^T$, from a normal distribution. The draws which imply autoregressive coefficients in the explosive region are rejected.

    \item \textit{Draw the stochastic volatilities}

    Obtain a draw of $\{\sigma_{\varepsilon}^{j}\}_{t=1}^T$ and $\{\sigma_{\eta_i}^{j}\}_{t=1}^T$ from $p(\{\sigma_{\varepsilon}\}_{t=1}^T|\phi^j,\{f_t^j\}_{t=1}^T)$ \\
    and from $p(\{\sigma_{\eta_i}\}_{t=1}^T|\bm{\Lambda}^j,\bm{\rho}^j,\{f_t^j\}_{t=1}^T,\mathbf{\tilde{y}})$, respectively.
    
    Draw the stochastic volatilities of the innovations to the factor and the idiosyncratic components independently, using the algorithm of \citet{kimstochvol1998}.

    \item \textit{Increase $j$ by 1 and iterate until convergence is achieved}
    
\end{enumerate}

We use 7,500 iterations and discard the first 2,500 to ensure convergence.

\subsection{A parsimonious model for computing recession probabilities}\label{recession_model}

Since the factor captures the business cycle dynamics of the overall economy, it can serve as input to a univariate regime-switching model to estimate recession probabilities. Following \citet{baumeister2024tracking}, we fit a Markov-switching model that distinguishes between heterogeneous recessions and expansions, allowing the depth and duration of each regime to vary over time.

A full characterization of regime dynamics in this context would require the development of a nonlinear dynamic factor model, as in \citet{chauvet2006dating} and more recently \citet{leiva2024real}. However, it is common practice to estimate regime-switching models in a second stage, using the output of a linear model as input. This approach also provides a way to evaluate the usefulness of our high-frequency index as an input for other statistical models.

We assume that the common factor $f_t$ can be decomposed into two components: a time-varying mean $\mu_t$ and a noise term $\epsilon_t$. It is assumed that there are two states of the economy identified by a latent discrete variable that equals 0 when the economy is in a recessionary episode, and 1 when the economy faces an expansionary episode, $s_t \in \{ 0,1 \}$. The time domain $t = 1,\ldots,T$ is partitioned into $N_0$ recessionary and $N_1$ expansionary
episodes, where a recession is followed by an expansion, which, in turn, must be followed by another recession. If the economy is in its $\tau_1$-th expansionary regime at time $t$, then $s_t = 1$ and the expected value of $f_t$ is equal to $\mu_{1,\tau_1}$, for $t\in \tau_1$. Instead, if the economy is in its $\tau_0$-th recessionary regime at time $t$, then $s_t = 0$ and the expected value of $f_t$ is equal to $\mu_{0,\tau_0}$, for $t \in \tau_0$. Formally, the factor move accord to the following dynamics:

\begin{align}
f_t=\mu_{0,\tau_0}(1-s_t)+\mu_{1,\tau_1}s_t+\epsilon_t, & & \epsilon_t\sim N(0,\sigma_{\epsilon}^2)
\end{align}

where $s_t$ is assumed to follow a two-state Markov chain defined by transition probabilities $P(s_t = i| s_{t-1} = j)=p_{i,j}$. Since there are two states, these probabilities can be summarized by the probability of remaining in expansion, $p$, and the probability of remaining in recession, $q$.

We assume Normal priors for $\mu_{0,\tau_0}$ and $\mu_{1,\tau_1}$, Beta priors for $p$ and $q$, and a Gamma prior for $\sigma_{\epsilon}^2$. See \citet{leiva2024real} for a detailed description of the model and Gibbs sampler.

\section{The High-Frequency Economic Index}\label{wei}

The factor time series, $f_t$, can be interpreted as a pseudo-weekly economic index. Following \citet{lewis2022measuring} and \citet{baumeister2024tracking}, we scale the index to match the mean and standard deviation of the year-over-year growth rate of real GDP. For the regime-switching model, we use the unscaled version of the index as input. We present results based on our preferred specification, which includes dynamic heterogeneity ($s = 1$) and stochastic volatility only in the factor innovations. Section~\ref{complexity} compares alternative model specifications and explains the rationale behind this choice.

A central question, however, is what the index is truly capturing. While much of the literature on diffusion indices evaluates performance through forecast accuracy, either for variables included in the model or for external targets, another perspective emphasizes the ability to summarize current macroeconomic conditions. This tradition, dating back to \citet{stock1989new}, relies on comparing the estimated index to reference variables or to officially defined business cycle phases, such as NBER recession dates.

In Ecuador, no official institution provides recession dating. Although the Central Bank of Ecuador publishes coincident and leading business cycle indices at a quarterly frequency, these are constructed from smoothed transformations of a set of macroeconomic variables and are designed to mimic the behavior of GDP rather than capturing the dynamics of the broader economy.\footnote{The methodology follows classical cycle approaches, often using filters such as Hodrick-Prescott or Christiano-Fitzgerald, which are less structurally grounded than dynamic factor models.} These indicators are released with considerable delays and are rarely cited by the Bank itself or by other public institutions, likely due to their lack of timeliness and the substantial ex post revisions introduced by the underlying statistical procedures as new data become available.\footnote{A related and legally codified notion is that of a "severe economic recession" (\textit{recesi\'on econ\'omica grave}), which is defined in Ecuadorian law as a period in which "the growth rate of Gross Domestic Product is negative for three consecutive quarters, either in quarter-over-quarter or year-over-year terms, or when the output gap is negative for two consecutive years." This condition is intended to serve as an exceptional clause that suspends certain fiscal rules. However, no official public reports have declared the existence of such a recession.}

As a consequence, economic downturns are typically labeled ex post using ad hoc or non-technical criteria. This institutional void complicates formal validation but also highlights the value of our contribution. To our knowledge, this is the first study to introduce a high-frequency, model-based framework for tracking Ecuador's business cycle in real time.

Given the absence of an official gold standard, we assess the plausibility of our index by examining its behavior during selected historical episodes and contrasting it with the dynamics of key macroeconomic variables.\footnote{Appendix \ref{app:comparison_BCE} presents a purely illustrative comparison of our index with business cycle estimates based on GDP reported with significant delay by the BCE. We emphasize that this comparison is not intended as formal validation, given the arguments presented earlier.} This approach allows us to characterize distinct phases of Ecuador's economic cycle in a systematic and replicable manner.

\begin{figure}
\centering
\caption{\label{fig:WEI}High-Frequency Economic Index}
\includegraphics[width=\textwidth]{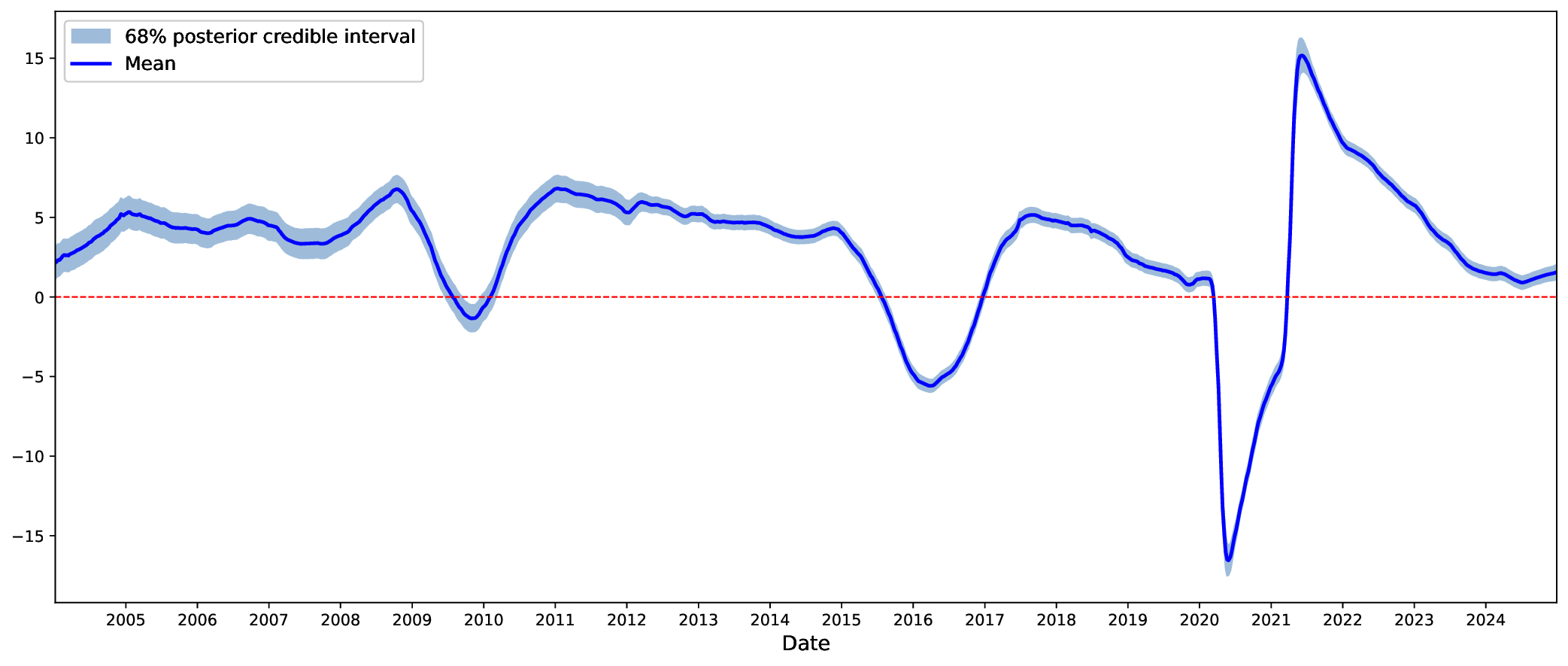}
\caption*{\tiny{\emph{Note:} The figure shows the mean of the posterior distribution of the High-Frequency Economic Index (HFEI), scaled to match year-over-year GDP growth following \citet{lewis2022measuring}. The shaded bands represent the 68\% posterior credible intervals.}}
\end{figure}

\begin{figure}
\centering
\caption{\label{fig:recprob}Recession Probability Over Time}
\includegraphics[width=\textwidth]{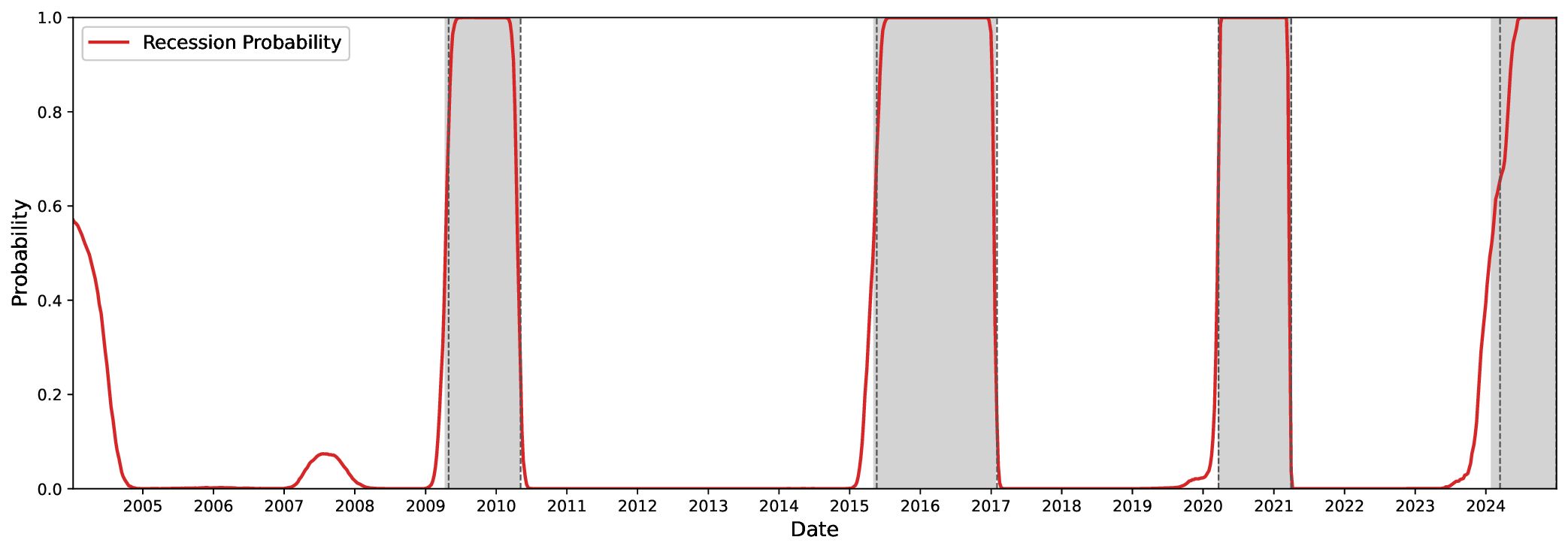}
\caption*{\tiny{\emph{Note:} The figure shows the estimated weekly probability of a recession. Shaded areas represent recession periods dated according to the Chauvet-Hamilton rule. Vertical dashed lines mark the moments when recessions and recoveries were called. A recession is called when the probability first exceeds 0.65, and dated to begin when the probability first crosses above 0.5. It ends when the probability drops below 0.35, and is dated to end when the probability first falls below 0.5.}}
\end{figure}

Figures~\ref{fig:WEI} and~\ref{fig:recprob} display the High-Frequency Economic Index and the mean of the estimated recession probabilities derived from the regime-switching model. Recession periods are highlighted using the rule proposed by \citet{chauvet2006dating}. The cyclical movements of the index closely mirror known expansions and contractions in the Ecuadorian economy. Moreover, the estimated recession probabilities provide a consistent and interpretable signal of the likelihood that the economy was in a recession at any given time.

A closer look at the index reveals a slight decline in activity associated with the 2008-09 collapse in oil prices. It also clearly captures the 2015-16 recession, which some policymakers incorrectly attributed to the earthquake that occurred in April 2016. As the index shows, the slowdown had already begun in early 2015, following the end of the oil boom in 2014.

The sharp contraction observed during the onset of the COVID-19 pandemic is also well represented, followed by a swift recovery in line with the gradual lifting of containment measures. Notably, the trough in economic activity is captured at the end of May, aligning with the final month in which mobility restrictions were broadly enforced. This timing is consistent with other high-frequency indicators, such as monthly national sales.

\begin{figure}
\centering
\caption{\label{fig:WEIvsGDP}Weekly Economic Index vs. Quarterly GDP}
\includegraphics[width=\textwidth]{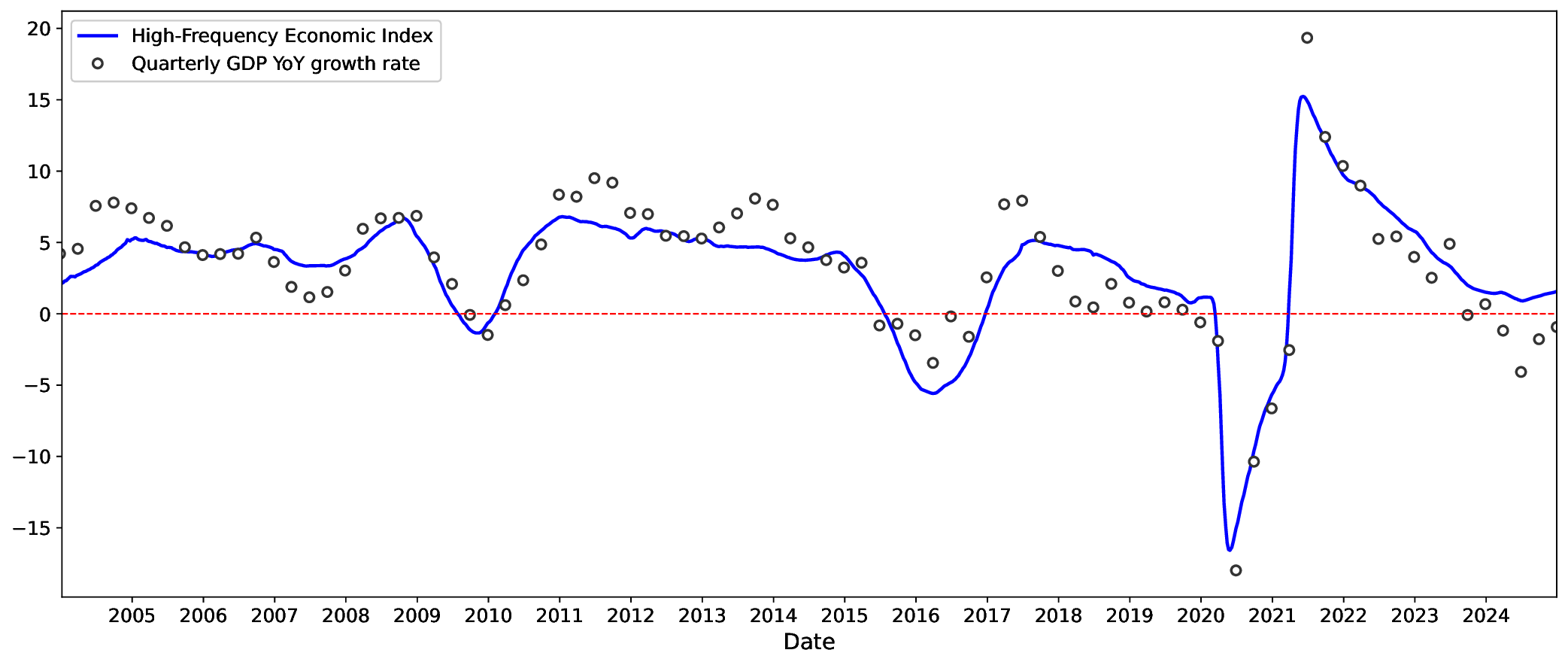}
\caption*{\tiny{\emph{Note:} The line shows the mean of the Weekly Economic Index (WEI), reported at weekly frequency. The black circles indicate actual quarterly year-on-year GDP growth rates. The GDP observations are plotted on the 28th of each quarter-ending month to align with the WEI timeline.}}
\end{figure}

Figure \ref{fig:WEIvsGDP} displays the HFEI alongside real GDP. Because the index is constructed from higher-frequency inputs, it provides substantially greater temporal resolution than GDP alone. It anticipates and mirrors the general direction of GDP movements while also identifying short-lived fluctuations that may be obscured in lower-frequency data. Periods of sustained growth are distinguishable from phases of deceleration or mild contraction, offering a timely and interpretable signal of underlying economic conditions.

Moreover, both the HFEI and the associated recession probabilities document four well-defined episodes of economic contraction. In contrast, GDP alone suggests more frequent ups and downs, and the business cycle indicators published by the Central Bank of Ecuador, which are based solely on GDP or constructed to maximize correlation with it, identify six contraction periods since 2004. However, this characterization is difficult to justify from a macroeconomic standpoint. For instance, during 2012-13, despite a mild deceleration in GDP growth, the annual rate never fell below 5\%, and no broad based deterioration in economic activity is observed. This illustrates the limitations of relying exclusively on GDP-driven signals. Our indicators, by contrast, are less prone to such false signals thanks to their multivariate foundation and the use of a regime-switching model with time-varying mean, which jointly enhance their ability to capture genuine turning points in real time.  Also, thanks to the inclusion of time-varying, regime-specific means, the model is able to distinguish recessions of varying depth as documented by \citet{leiva2024real}

An interesting feature of the HFEI is its behavior toward the end of the sample. The year 2024 was marked by weak economic activity, increased taxes, and a liquidity crunch in the public sector. However, government authorities failed to raise timely warnings about the deteriorating conditions. It was only by mid-2024 that the Central Bank of Ecuador acknowledged that the country had entered a recession at the start of the year. At the same time, it stated that the recession had already ended. Later, in early 2025, official figures for the third quarter of 2024, along with revisions to earlier GDP data, confirmed that the economy had contracted sharply in 2024.

One possible reason for the delayed recognition of the downturn is that traditional indicators, particularly credit and deposits, did not behave as they typically do during a contraction. Throughout 2024, these financial variables sent mixed signals. Most notably, deposits showed strong growth, with year-on-year rates reaching double digits. This unusual resilience in deposit behavior may have blurred the real-time assessment of the economy and contributed to the authorities' hesitation in acknowledging the recession.

\begin{figure}
\centering
\caption{\label{fig:HFEIvsfinancial}High-Frequency Economic Index vs. Deposits and Credit}
\includegraphics[width=\textwidth]{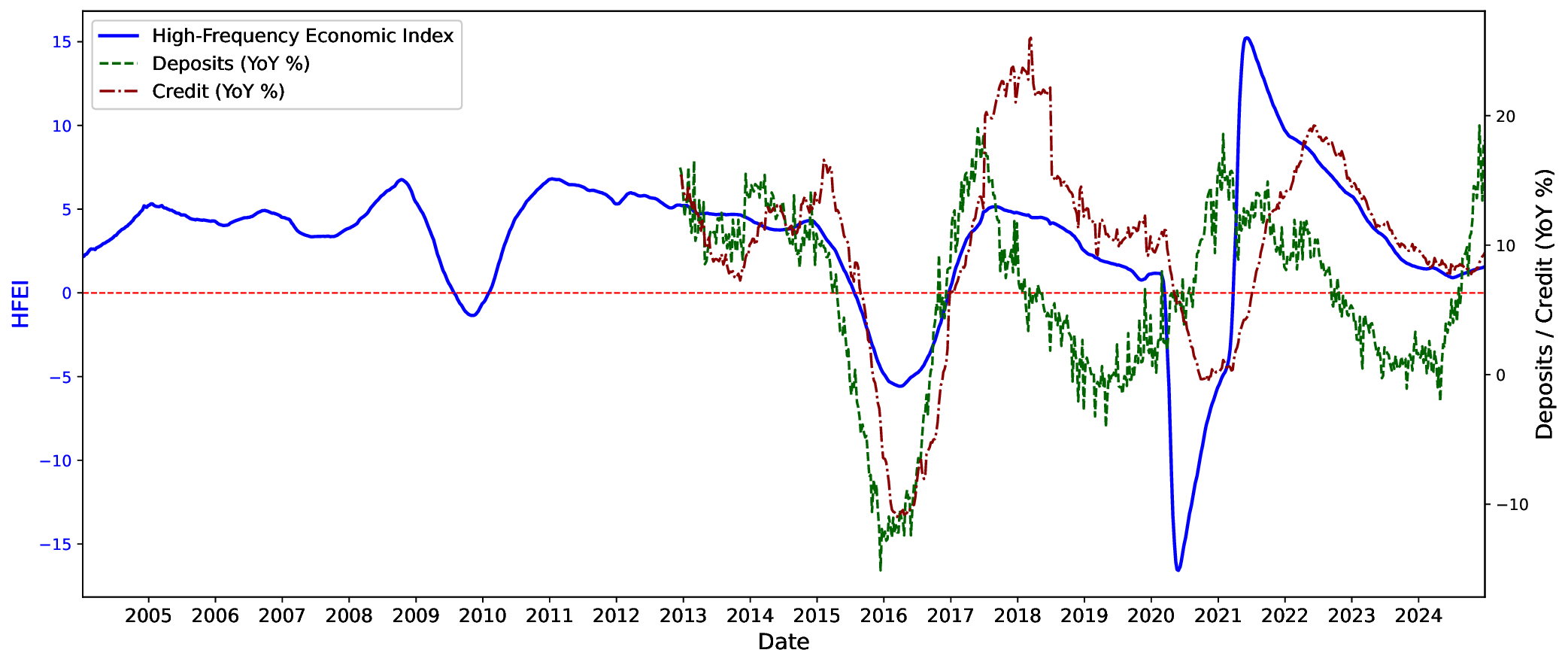}
\caption*{\tiny{\emph{Note:} The figure displays the High-Frequency Economic Index (HFEI) alongside year-on-year growth rates of deposits and credit to the private sector. All series are computed in pseudo-weekly frequency. The HFEI is shown on the left axis, while deposits and credit are plotted on the right axis.}}
\end{figure}

In fact, since the onset of the pandemic, monetary aggregates, particularly deposits, have exhibited behavior markedly different from that observed in previous business cycles. This shift is clearly visible in Figure~\ref{fig:HFEIvsfinancial}, which plots the HFEI alongside weekly measures of deposits and credit. The reasons behind this change are not entirely clear. Several structural developments after the pandemic may help explain it. For example, banking regulation underwent modifications, and some interest rate ceilings moved from fixed to floating formats. There was also a temporary increase in household savings due to confinement, similar to what was seen in other countries. In addition, passive interest rates, which define the return on savings deposits, increased, possibly encouraging deposit growth even in periods of weak real activity.

The High-Frequency Economic Index, by integrating information from a broad set of high-frequency indicators, captures this nuance. While it reflects a gradual slowdown over the course of 2024, the decline appears less abrupt than what is ultimately shown in the quarterly GDP data.

\begin{figure}
\centering
\caption{\label{fig:vol_WEI}Estimated stochastic volatility of the Weekly Economic Index}
\includegraphics[width=\textwidth]{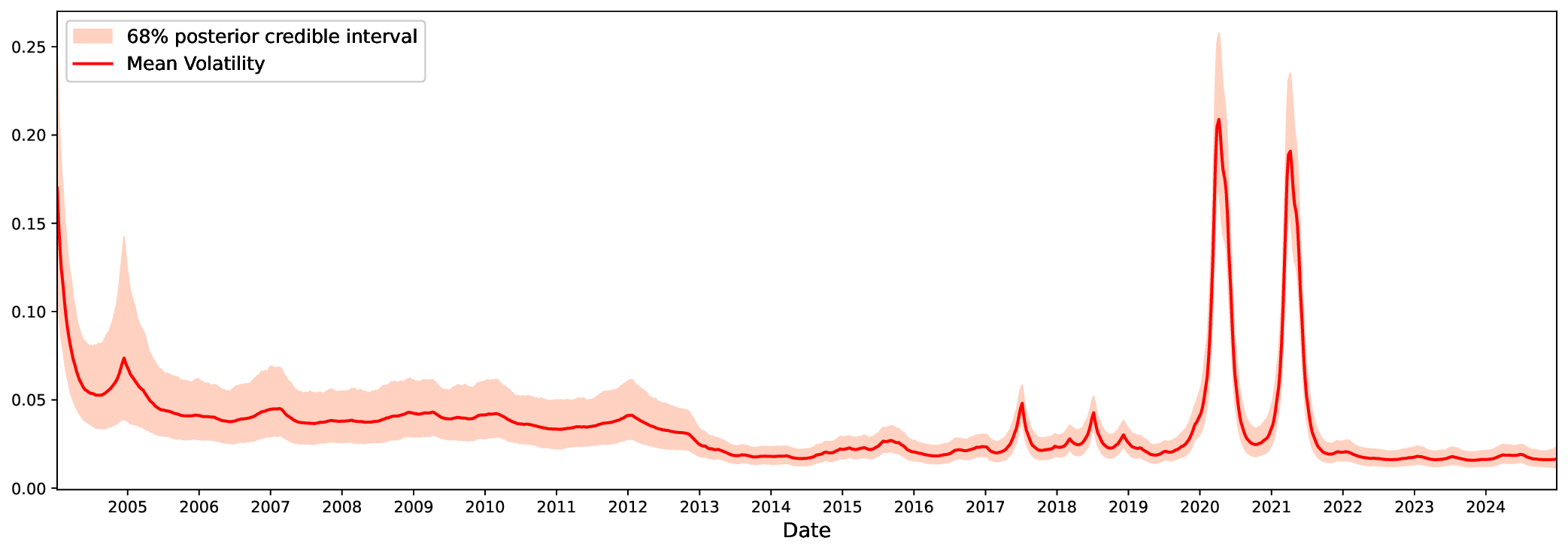}
\caption*{\tiny{\emph{Note:} This figure shows the estimated stochastic volatility of the factor underlying the Weekly Economic Index. The line represents the posterior mean of the volatility, and the shaded area denotes the 68\% posterior credible interval.}}
\end{figure}

Figure \ref{fig:vol_WEI} presents the posterior estimate of the stochastic volatility of the factor. The credible interval reveals substantial uncertainty during roughly the first half of the sample, primarily due to the relatively high number of missing observations at the beginning of the sample for some variables. Volatility tends to increase during periods of weak economic activity, a pattern also documented in studies such as \citet{antolin2024advances} and \citet{antolin2017tracking}.\footnote{\citet{antolin2017tracking} argue that this relationship between the first and second moments of the factor implies negative skewness in its distribution.} However, in our index high volatility is also observed during the peak of recovery periods, since the model is estimated using year-over-year growth rates. Dynamic factor models are traditionally estimated using month-over-month or quarter-over-quarter percentage changes, with the exception of GDP, which is typically entered in annualized terms. In such settings, volatility tends to capture both the downturn and the rebound within a short window. In contrast, using year-over-year growth rates causes elevated volatility to persist longer, particularly around turning points.\footnote{Related studies that adopt year-over-year growth rates in dynamic factor models include \citet{baumeister2024tracking} and \citet{gonzalez2019nowcasting}, although both assume homoskedastic conditional variances.}

\section{Does structural complexity pay off?}\label{complexity}

Table~\ref{tab:dic_comparison} reports the Deviance Information Criterion  for the eight model specifications considered. Two clear patterns emerge. First, the DIC strongly favors models with stochastic volatility, particularly those with volatility in the idiosyncratic errors. Second, the role of dynamic heterogeneity is less clear-cut: models with heterogeneous dynamics are preferred only when volatility is absent from the idiosyncratic component. These results suggest that allowing for time-varying volatility plays a more dominant role in improving in-sample fit than adding lagged factor loadings.

\begin{table}[htbp]
\centering
\small
\caption{Deviance Information Criterion (DIC)}
\label{tab:dic_comparison}
\begin{tabular}{lcc}
\toprule
\textbf{Stochastic Volatility} & \multicolumn{2}{c}{\textbf{Heterogeneous Dynamics}} \\
\cmidrule(lr){2-3}
 & No & Yes \\
\midrule
Factor and idiosyncratic errors & 12{,}223 & 12{,}301 \\
Idiosyncratic errors only       & 13{,}453 & 13{,}536 \\
Factor only                     & 14{,}418 & 14{,}026 \\
None                            & 15{,}282 & 15{,}147 \\
\bottomrule
\end{tabular}

\vspace{0.5em}
\begin{minipage}{0.95\textwidth}
\small \textit{Note:} Each row represents a different configuration of stochastic volatility. The columns indicate whether the model includes heterogeneous dynamics via lagged factor loadings. Lower DIC values indicate better in-sample fit.
\end{minipage}
\end{table}

While the DIC provides a useful summary of model fit adjusted for complexity, it is not the only relevant metric. We take the DIC as a guiding criterion, but also evaluate the behavior of the resulting economic indices to assess their plausibility and economic interpretability. This approach is in line with prior literature that uses DIC to discriminate between alternative dynamic factor models (e.g., \citealp{chan2016observed}), while still placing weight on the economic coherence of the latent factor.

\begin{figure}
\centering
\caption{\label{fig:WEI_models}High-Frequency Economic Index under Alternative Model Specifications}
\includegraphics[width=\textwidth]{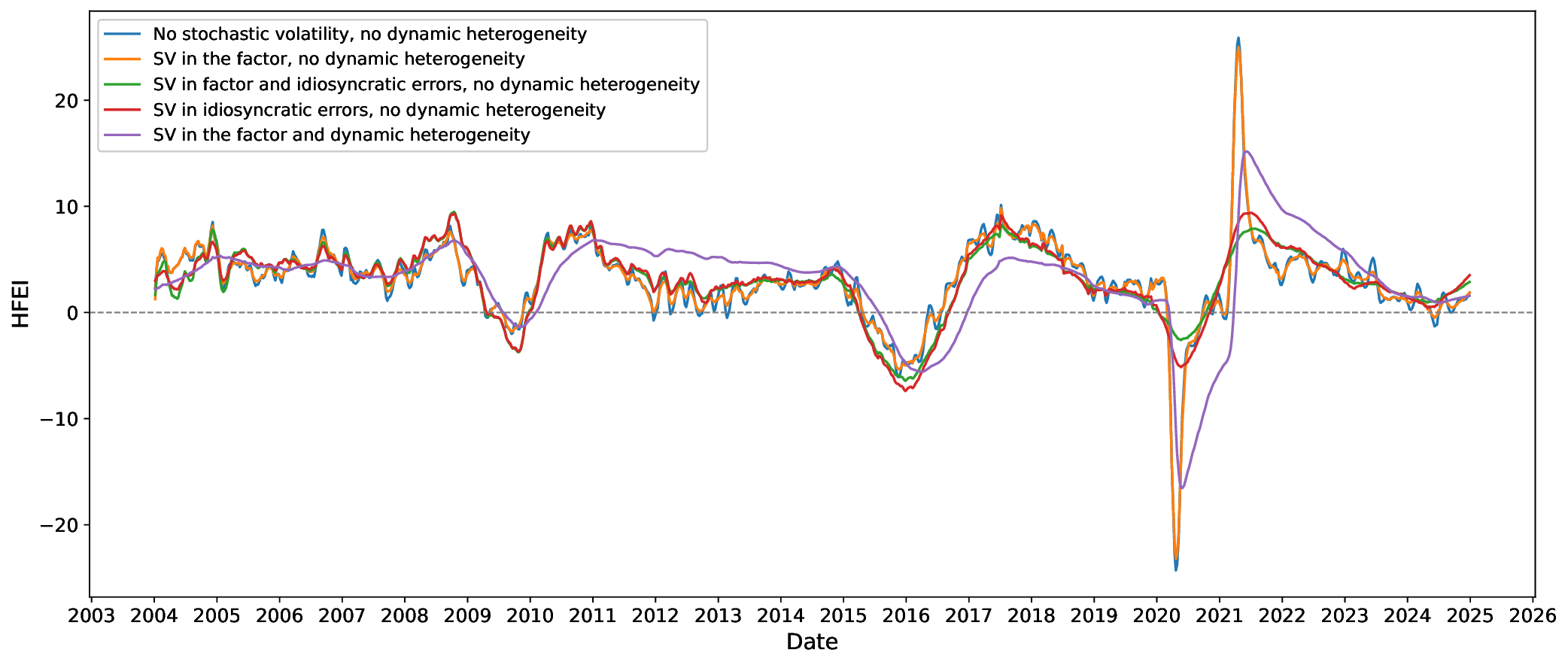}
\caption*{\tiny{\emph{Note:} This figure compares the estimated HFEI across five selected model configurations, varying in their treatment of stochastic volatility and dynamic heterogeneity. The index is scaled to match the mean and standard deviation of year-over-year GDP growth.}}
\end{figure}

Figure~\ref{fig:WEI_models} displays the High-Frequency Economic Index under five selected specifications. The remaining three are omitted for brevity, as their dynamics are largely redundant. Several insights emerge from these comparisons. Models with stochastic volatility in the idiosyncratic errors tend to produce smoother and more homogeneous cyclical movements. Under these specifications, the magnitude of the COVID-19 contraction appears surprisingly mild, comparable to or even smaller than prior recessions. This counterintuitive result likely reflects the model's ability to absorb abrupt variable-specific shocks through the volatility channel, thereby muting the response of the common factor. While this yields better in-sample fit, it may come at the cost of diminishing the factor's ability to represent aggregate economic conditions during extreme events.

In contrast, models without stochastic volatility, or those with stochastic volatility only in the factor but lacking dynamic heterogeneity, exhibit more jagged and noisy index behavior. These models require the factor to absorb more of the short-run fluctuations from volatile series, such as weekly trade indicators. This leads to greater volatility in the estimated factor, which may overstate transitory shocks.

The model with stochastic volatility in the factor and heterogeneous dynamics strikes a useful balance. Relative to its counterpart without dynamic heterogeneity, it exhibits smoother transitions and more clearly defined turning points. It also captures local peaks and troughs with greater flexibility, suggesting that dynamic heterogeneity allows the factor to incorporate information from series with asynchronous timing. This result is consistent with recent findings by \citet{antolin2024advances}, who show that when variables display delayed or leading relationships with the business cycle, standard models tend to underweight their contribution to the common factor. By contrast, including dynamic heterogeneity allows the model to exploit these temporal patterns and extract more balanced macroeconomic signals.\footnote{See \citet{antolin2024advances} for a more detailed discussion of dynamic heterogeneity in the context of nowcasting.}

Taken together, our findings suggest that structural complexity, particularly in the form of stochastic volatility, is rewarded in terms of statistical fit. However, capturing economically meaningful dynamics requires careful consideration of the tradeoff between model flexibility and interpretability. The specification that combines stochastic volatility in the factor with dynamic heterogeneity offers a compelling balance and is used as our baseline throughout the empirical analysis.

\section{Conclusions}\label{conclusions}

This paper develops a new High-Frequency Economic Index (HFEI) using a Bayesian dynamic factor model estimated on mixed-frequency data. By integrating official indicators published at quarterly, monthly, and weekly intervals, the model produces a timely and internally consistent summary of macroeconomic activity, helping fill critical informational gaps in data-constrained environments. We apply this approach to Ecuador, demonstrating its ability to track economic conditions in real time despite sparse and asynchronous data availability.

The proposed index accurately tracks major phases of Ecuador's business cycle, including global shocks like the 2008-09 financial crisis and the COVID-19 pandemic, as well as domestic episodes such as the 2015-16 oil-related downturn. Its real-time performance, particularly during the 2024 slowdown, underscores its practical relevance for short-term monitoring.

From a methodological perspective, the paper contributes to the literature by evaluating the trade-off between structural complexity and empirical performance across a grid of model specifications. We assess the roles of dynamic heterogeneity and stochastic volatility using the Deviance Information Criterion (DIC) and a narrative validation strategy. Our results suggest that incorporating stochastic volatility in the factor yields more interpretable and economically plausible estimates, even when not strictly favored by model selection criteria.

We also demonstrate how the HFEI can serve as an input to a second-stage Markov-switching model to estimate recession probabilities, offering a replicable, data-driven alternative to informal or retrospective recession dating practices in Ecuador.

The framework introduced here can be readily extended to other emerging markets. As access to high-frequency administrative and survey data expands, our approach provides a scalable blueprint for improving real-time macroeconomic analysis in data-scarce contexts.

\clearpage
\newpage
\appendix
\numberwithin{equation}{section}
\numberwithin{table}{section}
\numberwithin{figure}{section}
\section{Data Details}\label{app:data}

\subsection{Recovering adequate employment series from administrative data}

The National Institute of Statistics and Censuses (INEC) defines adequate employment as individuals who, during the reference week, receive labor income equal to or greater than the minimum wage and work at least 40 hours per week, regardless of their willingness or availability to work additional hours. This category also includes those who earn at least the minimum wage, work fewer than 40 hours, and do not wish to work more hours. The indicator is published monthly since September 2020. However, prior to that date, adequate employment was only reported each three months (March, June, September, and December), creating discontinuities in the time series.

To construct a continuous monthly series prior to September 2020, we use linear interpolation guided by administrative records from the Ecuadorian Social Security Institute (IESS), which are also published by INEC. These records report the number of formally affiliated workers and are available monthly from 2014 onward. Conceptually, both indicators aim to capture formal employment relationships. Adequate employment, as defined by INEC, includes individuals who meet minimum income and hour thresholds-criteria typically associated with formal jobs. Likewise, social security affiliation requires a formal labor contract and regular payroll contributions. While not identical, both measures reflect participation in the formal sector, and their co-movement over time suggests that affiliation records can effectively track changes in adequate employment. On average, approximately $78\%$ of adequately employed workers are affiliated with the social security system.

We estimate a linear regression using the available adequate employment observations, those recorded monthly since September 2020 and those reported for March, June, September, and December in earlier years, on monthly IESS affiliation data. The model includes a constant and a post-2019 dummy to capture potential structural changes stemming from modifications in the survey's sampling process. Using the estimated coefficients, we project the missing monthly values based on the social security data. This approach preserves the official figures and ensures smooth transitions between them, resulting in a consistent monthly series suitable for use in the model.

More sophisticated approaches could be used to address the missing values. For example, the original series with gaps could be incorporated directly into the main model, allowing the Kalman filter to infer the missing observations. It would also be feasible to extend the interpolation back to 2007, the year of the first available observation of adequate employment. However, we opt for a simpler strategy and restrict the interpolation to the period starting in 2014, as this reduces estimation uncertainty and preserves a clear conceptual relationship between adequate employment and its predictor.

We do not use the number of workers affiliated to the social security system directly in the model because this figure is published with more delay in comparison to adequate employment.

\subsection{Credits and deposits}

The weekly series for private sector credit and deposits were constructed from data published in the \textit{Bolet\'in Monetario Semanal} (BMS) of the Central Bank of Ecuador (BCE). This appendix outlines the steps followed to extract, clean, and transform the raw data into a consistent pseudo-weekly format.

The BMS publishes monetary aggregates at both monthly and weekly frequencies. Each weekly bulletin includes historical data in monthly frequency, along with weekly data for the most recent weeks. As a result, both frequencies coexist in a semi-structured layout. Although a consolidated database exists for monthly aggregates, no such dataset is available for weekly observations. For this reason, the historical weekly series were reconstructed through automated extraction and processing of individual bulletins published since 2012.

Several inconsistencies were identified in the extracted data, including mislabeling of dates in certain records. These issues were corrected by validating the reference dates of each observation. In some cases, erroneous entries were discarded, such as duplicates or entries whose dates conflicted with surrounding data points, to ensure continuity and chronological alignment of the series.

The available data were first organized in a pseudo-daily format. That is, as daily observations with irregular spacing. This structure allowed for consistent handling of reporting irregularities, such as gaps due to holidays or missing bulletins. The series were then aggregated into pseudo-weeks following the convention defined in Section \ref{pseudo-weeks}, in which each month is divided into four fixed intervals. For each pseudo-week, the average of the available daily observations was used to construct a representative weekly value.

Some pseudo-weeks had no available data due to publication omissions. To address these cases, missing values were imputed using information from the monthly series. When the last pseudo-week of a month was unavailable, the corresponding monthly value was assigned, ensuring internal consistency between the weekly and monthly versions of each series.  
After this procedure, a small number of missing pseudo-weekly values remained, 12 out of 607 observations for each series. These residual gaps were filled using the average of the preceding and following non-missing observations.

The resulting pseudo-weekly series were checked for continuity, alignment with monthly aggregates, and absence of artificial jumps introduced by the imputation process. The final dataset preserves the original scale of the variables (millions of U.S. dollars) and reflects the best-available weekly signal derived from the BMS publications.

\subsection{Trade variables}

The pseudo-weekly series for exports and imports were constructed from transaction-level customs data published by the BCE. This appendix outlines the steps followed to extract, classify, and aggregate the raw data into a consistent pseudo-weekly format.

The underlying data were obtained from the BCE's Foreign Trade Information Platform, which allows queries by month and year. However, the underlying database enables the identification of transaction-level customs records disaggregated by tariff subheading and date of transaction. This, in turn, allows for the construction of a high-frequency dataset and the classification of trade flows into categories that may behave differently across distinct phases of the business cycle.
 
A major limitation was identified in the export records prior to 2012, related to the migration from the SICE system to ECUAPASS.\footnote{SICE (Interactive Foreign Trade System) was the platform used by Ecuador's customs authority until 2011 to record foreign trade operations. In 2012, it was replaced by ECUAPASS, an automated system that enables more accurate and continuous management of customs declarations.} Before this transition, many transactions were concentrated at the beginning of each month, likely due to limitations in how transactions were recorded or processed. This pattern introduces artificial spikes and distorts the temporal distribution of trade flows. To address this issue, we restrict the export sample to the period from 2012 onward. Import records do not exhibit the same distortion and are used from 2004 onward, consistent with the rest of the dataset.
 
The raw customs records were grouped into analytical categories for aggregation. Export data were classified into oil and non-oil categories, following the official methodology of the BCE. Import data were divided into consumer goods, capital goods, raw materials, and other goods, although only the first three are used in the model. These groupings are based on tariff subheading codes and required harmonization across different versions of the Harmonized System (HS), which has been revised periodically over time.

Days with no recorded transactions were treated as genuine zero flows rather than missing data and explicitly retained in the dataset. The cleaned daily series were then aggregated into pseudo-weeks using the convention defined in Section~\ref{pseudo-weeks}, summing the daily observations within each interval to construct consistent weekly flow variables.

The final pseudo-weekly series are consistent in terms of measurement, aggregation, and classification. They serve as a high-frequency signal of external demand and supply conditions and are directly integrated into the dynamic factor model.

\section{A closer look at the BDFM}\label{app:model}
The model in state-space form is given by:

\begin{align}
\mathbf{\tilde{y}}_{t} &= \mathbf{H}\bm{\xi}_{t}\label{app_H}\\
\bm{\xi}_{t}& =\mathbf{F}\bm{\xi}_{t-1}+\mathbf{R}_{t}\mathbf{v}_{t}\label{app_FR}\\
\mathbf{v}_{t}&\sim N(\mathbf{0},\mathbf{I})
\end{align}

where $\mathbf{\tilde{y}}_{t}$ is the vector of observed variables. We order the quarterly variables at the beginning of $\tilde{\mathbf{y}}_{t}$, then the monthly variables and finally the weekly ones.

The state vector $\mathbf{\xi}_{t}$ is defined as:\footnote{Here, we assume $s=0$, imposing no dynamic heterogeneity in deriving the system matrices. For $s=1$, the state vector includes one additional lag of the factor, and the system matrices are adjusted accordingly.}

\[
\mathbf{\xi}_{t}=\left[f_{t},\ldots,f_{t-11},\mathbf{u}^q_1,\ldots,\mathbf{u}^q_{n_q},\mathbf{u}^m_1,\ldots,\mathbf{u}^m_{n_m},\mathbf{u}^w_1,\ldots,\mathbf{u}^w_{n_w}\right]^{'}
\]

\[
\mathbf{u}_i^q=\left[u_{i,t}^q,u_{i,t-1}^q,\ldots,u_{i,t-11}^q\right]
\]

\[
\mathbf{u}_j^m=\left[u_{j,t}^m,u_{j,t-1}^m,\ldots,u_{j,t-d}^m\right]
\]

\[
\mathbf{u}_k^w=\left[u_{k,t}^w,u_{k,t-1}^w,\ldots,u_{k,t-p_q}^w\right]
\]

where $d$ is the maximum of $p_q$ and 4. The dimension of the state vector is $n_{s}\times1$ where $n_{s} = 12+12n_q+(d+1)n_m+(q+1)n_w$. Note that this definition of the state vector, along with the composition of the matrices that we will explain later, makes it clear that we assume $p_f$ and $p_q$ are always less than 12, an assumption that is plausible for any economy.

Let $\bm{\Lambda}^q$, $\bm{\Lambda}^m$ and $\bm{\Lambda}^w$ be the vectors of loadings. Let $\bm{\phi}=[\phi_1,\phi_2,\cdots,\phi_{p_f}]$ be the vector of autoregressive coefficients of the factor. Let $\bm{\rho}_i^q=[\rho_{i,1}^q,\rho_{i,2}^q,\cdots,\rho_{i,p_q}^q]$, $\bm{\rho}_j^m=[\rho_{j,1}^m,\rho_{j,2}^m,\cdots,\rho_{j,p_q}^m]$ and $\bm{\rho}_k^w=[\rho_{k,1}^w,\rho_{k,2}^w,\cdots,\rho_{k,p_q}^w]$ the vector of autoregressive coefficients for the $i$-th quarterly variable, the $j$-th monthly variable and the $k$-th weekly variables. Let $\bm{\sigma}^q_t=[\sigma^q_{\eta_1,t},\cdots,\sigma^q_{\eta_{n_q,t}}]'$, $\bm{\sigma}^m_t=[\sigma^m_{\eta_1,t},\cdots,\sigma^m_{\eta_{n_m,t}}]'$ and $\bm{\sigma}^w_t=[\sigma^w_{\eta_1,t},\cdots,\sigma^w_{\eta_{n_w,t}}]'$ be the vector of innovation standard deviations.

The matrix $\mathbf{H}$ in \ref{app_H} is:

\[
\mathbf{H}=\begin{bmatrix}
\mathbf{H}^q \\
\mathbf{H}^m \\
\mathbf{H}^w \end{bmatrix}
\]

where $\mathbf{H}^q$, $\mathbf{H}^m$ and $\mathbf{H}^w$ are given by:

\[
\mathbf{H}^{q}=\begin{bmatrix} \frac{1}{12}\operatorname{diag}(\mathbf{\Lambda}^{q})\mathbf{1}_{[n_q\times12]} &  \frac{1}{12}\mathbf{I}_{n_q}\otimes \mathbf{1}_{[1\times12]} & \mathbf{0}_{[n_q\times (n_m(d+1))]} &  \mathbf{0}_{[n_q\times(n_w(p_q+1))]}
\end{bmatrix}
\]

\[
\mathbf{H}^{m}=\begin{bmatrix} \frac{1}{4}\operatorname{diag}(\mathbf{\Lambda}^{m})\mathbf{1}_{[n_m\times4]} & \mathbf{0}_{[n_m\times(8+12n_q)]} & \frac{1}{4}\mathbf{I}_{n_m}\otimes \mathbf{1}_{[1\times(d+1)]} &  \mathbf{0}_{[n_m\times(n_w(p_q+1))]}
\end{bmatrix}
\]

\[
\mathbf{H}^{w}=\begin{bmatrix} \mathbf{\Lambda}^{w} & \mathbf{0}_{[n_w\times(12+12n_q+n_m(d+1)-1)]} & \mathbf{I}_{n_w}\otimes [1, \mathbf{0}_{[1\times p_q]}]
\end{bmatrix}
\]

The matrix $\mathbf{F}$ in \ref{app_FR} is:

\[
\mathbf{F}=\begin{bmatrix} \mathbf{F}^{f} &  &  & \mathbf{[0]} \\
 & \mathbf{F}^q & &  \\
 &  & \mathbf{F}^m &  \\
\mathbf{[0]} &  &  & \mathbf{F}^w
\end{bmatrix}
\]

where:

\[
\mathbf{F}^f = \begin{bmatrix}
\phi_1 & \phi_2 & \dots & \phi_{p_f} & \mathbf{0}_{[1\times(12-p_f)]} \\
 &  &  &  &   \\
 &  & \mathbf{I}_{[11]} & & \mathbf{0}_{[11\times1]} \\
 &  &  &  &  & 
\end{bmatrix},
\]

\[
\mathbf{F}^q=\begin{bmatrix} \mathbf{F}_1^q &  & \mathbf{[0]} \\
 & \ddots &  \\
\mathbf{[0]} &  & \mathbf{F}_{n_q}^q
\end{bmatrix},
\quad
\mathbf{F}^m=\begin{bmatrix} \mathbf{F}_1^m &  & \mathbf{[0]} \\
 & \ddots &  \\
\mathbf{[0]} &  & \mathbf{F}_{n_m}^m
\end{bmatrix},
\quad
\mathbf{F}^w=\begin{bmatrix} \mathbf{F}_1^w &  & \mathbf{[0]} \\
 & \ddots &  \\
\mathbf{[0]} &  & \mathbf{F}_{n_w}^w
\end{bmatrix}
\]

For each $i=1,\cdots,n_q$, $j=1,\cdots,n_m$ and $k=1,\cdots,n_w$, the matrices in $\mathbf{F}^q$, $\mathbf{F}^m$ and $\mathbf{F}^w$ are defined as:

\[
\mathbf{F}_i^q=\begin{bmatrix} \rho_{i,1}^q & \rho_{i,2}^q & \cdots & \rho_{i,p_q}^q & \mathbf{0}'_{[12-p_q]}\\
 1 & 0 & \cdots &  0 & \mathbf{0}'_{[12-p_q]}\\
 \vdots & \vdots & \ddots & \vdots & \vdots \\
0 & 0 & \cdots & 1 & \mathbf{0}'_{[12-p_q]}
\end{bmatrix}
\]

\[
\mathbf{F}_j^m=\begin{bmatrix} \rho_{j,1}^m & \rho_{j,2}^m & \cdots & \rho_{j,p_q}^m & \mathbf{0}'_{[d+1-p_q]}\\
 1 & 0 & \cdots &  0 & \mathbf{0}'_{[d+1-p_q]}\\
 \vdots & \vdots & \ddots & \vdots & \vdots \\
0 & 0 & \cdots & 1 & \mathbf{0}'_{[d+1-p_q]}
\end{bmatrix}
\]

\[
\mathbf{F}_k^w=\begin{bmatrix} \rho_{k,1}^w & \rho_{k,2}^w & \cdots & \rho_{k,p_q}^w & 0\\
 1 & 0 & \cdots &  0 & 0\\
 \vdots & \vdots & \ddots & \vdots & \vdots\\
0 & 0 & \cdots & 1 & 0
\end{bmatrix}
\]

Finally, $\mathbf{R}_{t}$ is time-varying because of the stochastic volatilities.

\[
\mathbf{R}_t=\operatorname{diag}\begin{bmatrix} \sigma_{\varepsilon_t}\\
 (\mathbf{I}_{n_q}\otimes \mathbf{e}_1^{12})\bm{\sigma}^q_t \\
 (\mathbf{I}_{n_m}\otimes \mathbf{e}_1^{d+1})\bm{\sigma}^m_t \\
 (\mathbf{I}_{n_w}\otimes \mathbf{e}_1^{p_q+1})\bm{\sigma}^w_t
\end{bmatrix}
\]

where $\mathbf{e}_1^{N}$ is a vector of size $N\times 1$ with a 1 in the first entry and 0 elsewhere.

\section{A Brief Comparison with BCE Estimates of the Business Cycle}\label{app:comparison_BCE}

\begin{figure}
\centering
\caption{\label{fig:HFEIvsCycle}High-Frequency Economic Index and BCE Business Cycle Estimates}
\includegraphics[width=\textwidth]{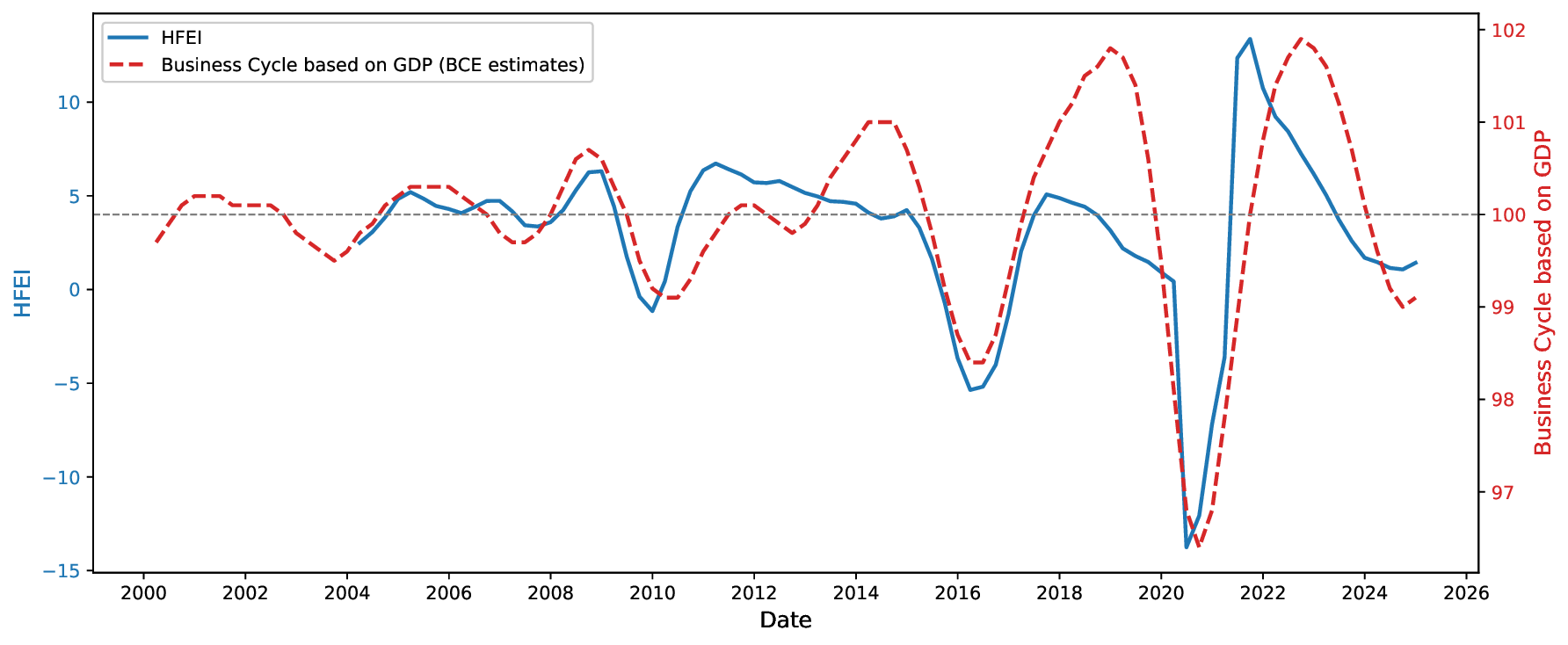}
\end{figure}

Figure \ref{fig:HFEIvsCycle} compares our High-Frequency Economic Index (HFEI) with the business cycle estimates published by the BCE based on real GDP. Although the long-term patterns of both indices appear broadly similar, with a simple correlation coefficient of approximately 0.7, important differences emerge upon closer inspection.

The BCE's GDP-based business cycle estimates exhibit a highly regular cyclical pattern, with expansions and contractions occurring at fairly uniform intervals and with peaks and troughs of similar magnitudes. For instance, the pre-pandemic expansion appears roughly as strong as the post-pandemic recovery, a result that is difficult to justify. This rigid behavior largely reflects the methodology used to extract cyclical components and the limitations of relying on GDP as the sole measure of economic activity.

While the BCE also publishes coincident indicators, these are explicitly constructed to maximize correlation with the GDP cycle, thereby offering limited independent information. Our HFEI, in contrast, leverages mixed-frequency data and aims to capture more nuanced and timely signals of economic dynamics beyond those embedded solely in GDP.

\clearpage

{\small{}\bibliographystyle{ecta-fullname}

}{\small \par}
\end{document}